# The Design of Programmable Transmitarray with Independent Controls of Transmission Amplitude and Phase

He Li, Yun Bo Li, *Member, IEEE*, Chao Yue Gong, Shu Yue Dong, Shi Yu Wang, Hai Peng Wang, and Tie Jun Cui, *Fellow, IEEE*

*Abstract*—A design method of programmable transmitarray with independent controls of transmission amplitude and phase is proposed in C-band. The unit cell with cascaded structures mainly consists of four parts, including the receiving antenna, reconfigurable attenuator with PIN diodes, reconfigurable phase shifter with varactors and transmitting antenna. Correspondingly, various manipulations of spatial electromagnetic (EM) fields are achieved by varying the bias voltages of PIN diodes and varactors in the transmitarray. The fabricated unit cell is measured in a standard waveguide, and the whole array with 8×8 unit cells is measured with two horns for calibration of the programmable EM features. The experimental results show that the transmission magnitude can range from -16 dB to -3.6 dB and the transmission phase achieves 270º coverage independently under the 16-bit programmable control. To further exhibit the capability and functionality of the proposed transmitarray, the waveform engineering of adaptive beamforming and power- allocation beamforming are separately realized in the experiment. The measured results have good agreements with our theoretical calculations, verifying the validity of our design method.

## I. INTRODUCTION

TRANSMITARRAYs [1-6], as the new architectures for implementing high-directivity beamforming aperture, have received much attention in recent years. With the advantages of low profile, easy fabrication, and high efficiency, the transmit-arrays have big potentials in advanced wireless communication systems and the radar systems. Compared with reflectarrays [7-9] that are suffered from the effect of feed blockage and the active phase array with expensive transmitting and receiving modules, they are becoming more attractive for applications in waveform engineering. The transmitarrays can be regarded as the electromagnetic (EM) transmission metasurfaces composed of period subwavelength unit cells. By controlling the phase shift and amplitude of unit cell, the metasurfaces have powerful abilities to manipulate the incident EM waves.

The passive transmission metasurfaces [10-14] have been researched over much of the spectrum, and rapidly developed towards low profile, broadband and multiple functionalities. Under the premise of the required full phase coverage, some efforts are made to realize ultra-thin structure with less number of layers by using slot antenna [11] or creating compact Hygens' structure [12]. In addition, planar gradient refractive index metasurface lens [15] with impendence matching method has been well researched for its characteristics of high efficiency and broader bandwidth compared with the transmitarrays recently. However, this type of lens [16] is always composed of two impendence matching parts and a core part with massive printed circuit board (PCB) layers, which increase the thickness of the lens antenna. And it can be considered that the wider bandwidth is achieved by sacrificing the profile.

Compared with the passive transmitarrays which have the single function, active or programmable versions can generate dynamic apertures to control the spatial EM wave in real time by integrating some active elements such as PIN diodes [1, 5, 17] , varactor diodes [18] , MEMS [2, 3] and liquid crystal [19] into unit cells. Generally, the active design methods consist of three types: electric resonance [6], Hygens' [20, 21] and Rx-phase shifter-Tx [1, 2]. Considering of thick design under the electric resonance approach and achievement of ultra-narrow bandwidth based on the Hygens' method, the Rx-programmable phase shifter-Tx method is mostly preferred to actualize the work of beam scanning in nowadays with the performance of low-profile and wide-band. This method innovatively builds a bridge between the spatial EM field and guided EM wave, in which the receiving and transmitting antennas are regarded as the spatial wave input and output devices respectively, and the reconfigurable phase shifter is considered as the programmable information processor of the guided wave to achieve full phase coverage with high transmission efficiency. Accordingly, the design of active beam steering based on the above method is proposed. Further, the active transmitarrays of the wideband [22-24] and polarized controls [18] are also presented. Also, the information processor can not only be implemented as the reconfigurable phase shifter, but the amplification [25] and nonreciprocal

This work was supported in part by National Natural Science Foundation of China (61901113, 61801262, 61571117, 61631007); in part by Foundation Strengthening Program Technology Field Fund (2020-JCJQ-JJ-266) and in part by General Technical Research Project (20201116-0155-001-001).

The authors are with the State Key Laboratory of Millimeter Waves, Southeast University, Nanjing 210096, China (e-mail: mozartliyunbo@163.com, tjcui@seu.edu.cn).

circuits [26] by introducing the power amplifiers.

Most of the previous researchers only focus on the phase reconfiguration in design of metasurfaces. However, more functional and required applications cannot be achieved with the lack of amplitude information. Thus, controlling the phase and amplitude to complete the EM beams in the far-field has been presented although it suffers from hard design progress and high profile based on the electric resonance method [27]. An alternative strategy [28], using the combination of C-ring and rod textures to controls the amplitude and phase under cross polarization, can have the good performances of low profile and easy fabrication. Later, to simplify the above design, only C-ring-like structures are used with lack of the rod [29-32]. In addition, the approach of combining the amplitude-control and phase-control metasurfaces [33] was designed to suppress the sidelobe levels. The two metasurfaces are cascaded with the large distance for neglecting the sample effect, which increases the thickness of the whole transmitarray. Also, the reflected EM waves of the amplitude-control metasurface will increase the back-lobe levels. All of above publications are concentrated on the passive designs with fixed EM functions.

To dynamically manipulate the amplitude and phase of the unit cell independently in real time, here we introduce cascaded programmable attenuator and phase shifter embedded with the active elements in the frame of Rx-programmable information processor-Tx type transmitarray. Further, in order to verify the validity of the proposed design, adaptive beamforming [34, 35] which can realize both of main beams and null points is achieved with the algorithm of maximum signal-to-interference ratio (SIR) [36], and multi-beams with the required power allocation [27] is generated by the programmable transmitarray. Compared with the control of phase only in our experiment, better results are observed in achieving adaptive beamforming and multi-beams with power allocation using programmable controls of the amplitude and phase independently.

The paper is organized as follows. The architecture of the proposed unit cell in transmitarray is introduced in Section II. The theoretical analyses of beamforming based on the adaptive and power allocation algorithms are presented in Section III. Finally, the experimental calibration of the fabricated prototype is completed and the corresponding beamforming results of the programmable transmitarray are shown in Section IV.

## II. Design of Unit Cell in Transmitarray

The unit cell of the programmable transmitarray with independent controls of the transmission amplitude and phase is shown in Fig. 1. The whole unit cell of the transmitarray mainly consists of the cascade textures including the receiving antenna, programmable reflection-type attenuator and phase shifter, and transmitting antenna. To realize the receiving and transmitting of spatial EM fields, two pairs of stacked patch antennas with the same sizes are preferred. The bow-tie like and squared patch antennas are cascaded with air space to increase the working bandwidth in our design. The bowtie antenna is located on the substrate F4B220 with the dielectric constant of 2.2 and the dielectric loss tangent of 0.003. In addition, the substrates of programmable attenuator, programmable phase shifter and patch antennas are chosen as F4B265 with the dielectric constant of 2.65 and the dielectric loss tangent of 0.003. And the adhesive coating of Rogers 4450f with dielectric constant of 4.55 and the dielectric loss tangent of 0.0037, is selected to combine the top and bottom structures in the middle layer.

The core designs to achieve the dynamic controls of transmission amplitude and phase independently are the programmable attenuator and phase shifter. At first, the two active components should be tailored separately, and the cascaded structures combined with the two pairs of passive stacked patches should be optimized together.

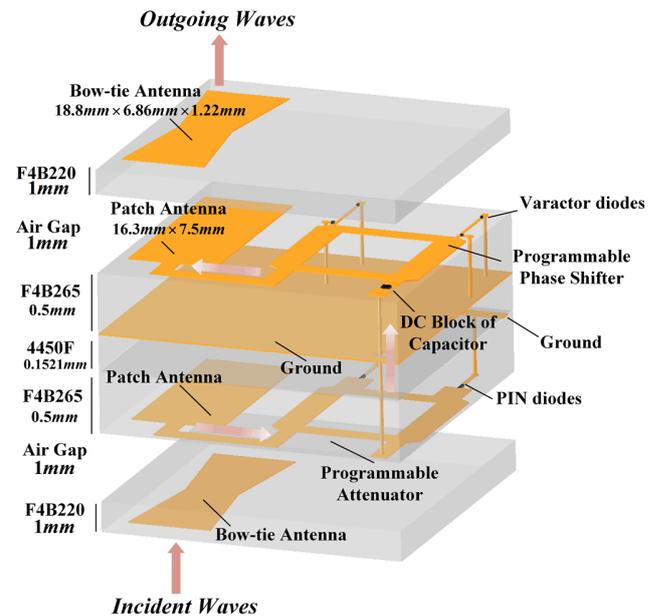

Fig. 1 The schematic diagram of unit cell in proposed transmitarray.

### A. Programmable Attenuator

The control of transmission amplitude can be realized using the active design of programmable attenuator with PIN diodes. And also, the high reflection coefficients caused by the low transmission amplitude need to be avoided while changing the bias voltages of the active elements. Correspondingly, the reflection-type low-phase-shift programmable attenuator [37, 38] with PIN diodes is preferred, as shown in Fig. 2. This reflection-type programmable attenuator mainly consists of the two parts of the 3 dB branch-line coupler and two identical short circuits. In our design, each short circuit is composed of the PIN diode and the transmission line working as the phase compensation circuit. The guided waves can arrive at the two identical short circuits with the same amplitude and 90-degree phase difference. According to the EM characteristics of the 3 dB coupler, the energies that are reflected by the metal via connecting with the ground, are combined with the same phase shift at the output port. The reflected EM waves at the input port are cancelled by composition with inverted phase, and the low reflection coefficients can be obtained accordingly.

In general, the PIN diodes are applied with the two conditions of ON and OFF, referring to the short and the open states respectively, which can also be considered as the 0 ohm and ∞ ohm in circuits. The PIN diodes can be generally modeled with the equivalent circuit shown in Fig 2(b), in which the *R, L, C* are the constant parameters that are dependent on the geometry of the intrinsic layer in the PIN diodes. And the resistance *RR* is varied when the voltage bias of the diode changed, that can be introduced into the programmable attenuator for the dynamic control of transmission amplitude. In our design, the *Skyworks SMP1320-079LF* is chosen and integrated on the circuit. The detailed parasitic parameters of this PIN diode are described in Fig 2, and the *RR* can be tuned from 0 ohm to ∞ohm while varying the voltage bias ideally, which is extracted in the commercial software *ADS 2014*.

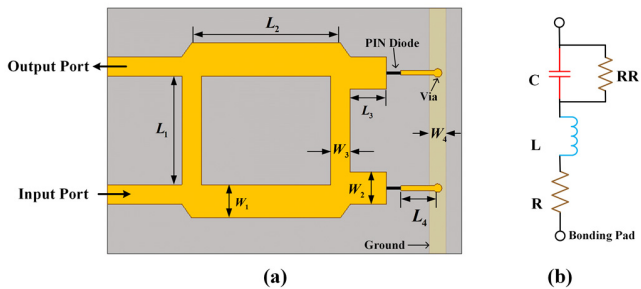

Fig. 2 Implementation of the reflection-type low-phase-shift programmable attenuator. (a) The schematic of the programmable attenuator. (b). The equivalent model of PIN Diode which is Skyworks SMP1320-079LF (C = 0.23pF, L = 0.7nH, R = 0.3Ω).

To dynamically manipulate the transmission amplitude with low-phase-shift, the phase compensating network needs to be introduced, which also works as the impendence matching circuit. Taking the equivalent model of the PIN diodes as the RLC boundaries into the commercial software *HFSS 2018*, the transmission coefficient and phase shift between the input port and the output port are both simulated. Further, the sizes of the programmable attenuator are optimized in the unit cell to achieve high performance. And the final detailed size parameters of the 3 dB branch-line coupler and the short circuits in programmable attenuator are shown in Table I. As a result, the transmission amplitude of unit cell can be manipulated realized by proposed programmable attenuator with low-phase-shift.

### B. Programmable Phase Shifter

As another critical parts in designing the unit cell, the programmable phase shifter generally is designed to achieve large phase coverage by changing the bias voltage on the active elements. And the good performances of insert loss and uniformity are required at the same time. The most common approach is the reflection-type programmable phase shifter [39-42] which usually employs the 3 dB branch line coupler and the reflective networks containing the active components. The corresponding design is shown in Fig. 3(a), which is similar to the proposed programmable attenuator in the previous part.

To construct the reflection-type programmable phase shifter, two identical π-type reflective loads with four varactors are synthesized in our design. Each π-type reflective network consists of a transmission line in parallel with two varactors at both ends. The *MACOM MA46H120* varactors are selected in this circuit. Correspondingly, the total capacitance of the varactors can range from 0.2 pF to 0.8 pF composing of a parasitic series resistance of 2 ohm with changing the reverse bias voltage. The feature of the varactors allows the control of phase variation to be continuous. By taking the equivalent circuit as the RLC boundaries into the commercial software *HFSS 2018*, the programmable phase shifter is simulated and analyzed in the whole texture of the transmitarray unit cell.

Due to the characteristics of 3 dB branch-line coupler, the reflection-type programmable phase shifter can achieve the good matching at the input and output ports, that extremely decreases the return loss. In order to achieve the large phase shift and small variation of the transmission coefficients at the same time, the detailed parameters are optimized which are shown in Table I.

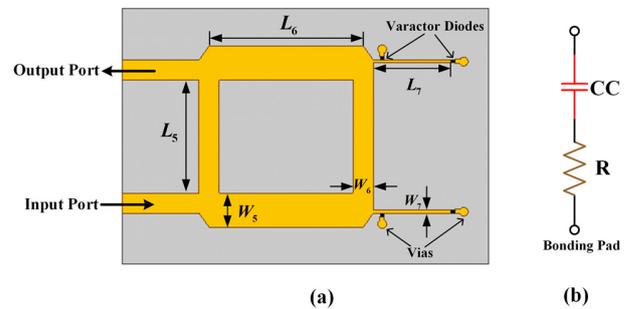

Fig. 3 Implementation of the reflection-type programmable phase shifter. (a) The schematic of the programmable phase shifter. (b). The equivalent model of varactor which is MACOM MA46H120 (R = 2Ω).

TABLE I
DIMENSIONS OF TRANSMITARRAY UNIT CELL

| Parameter | Value(mm) | Parameter | Value(mm) |
|---|---|---|---|
| $L_1$ | 7.74 | $W_1$ | 2.35 |
| $L_2$ | 10.5 | $W_2$ | 2.3 |
| $L_3$ | 2.57 | $W_3$ | 1.38 |
| $L_4$ | 2.6 | $W_4$ | 1.2 |
| $L_5$ | 7.74 | $W_5$ | 2.35 |
| $L_6$ | 10.5 | $W_6$ | 1.38 |
| $L_7$ | 5.3 | $W_7$ | 0.2 |

### C. Full-Wave Simulation Results

By cascading the receiving antenna, the programmable attenuator, the programmable phase shifter and the transmitting antenna described in Fig. 1, the programmable transmitarray unit cell can be simulated and optimized entirely to achieve the independent controls of transmission amplitude and phase. The dimension of the unit cell is $25mm \times 25mm$, corresponding to $0.458\lambda \times 0.458\lambda$ at the frequency of 5.5 GHz. The full-wave simulation of the unit cell is implemented in the commercial software *HFSS 2018* under the boundary condition of period and the excitation of Floquet ports. And the equivalent circuits of the proposed PIN diodes and varactors are loaded as the RLC boundaries in our simulation. To illustrate the ability of independent control, the programmable attenuator and the phase shifter are tailored with varying the resistance *RR* and capacitance *CC* respectively.

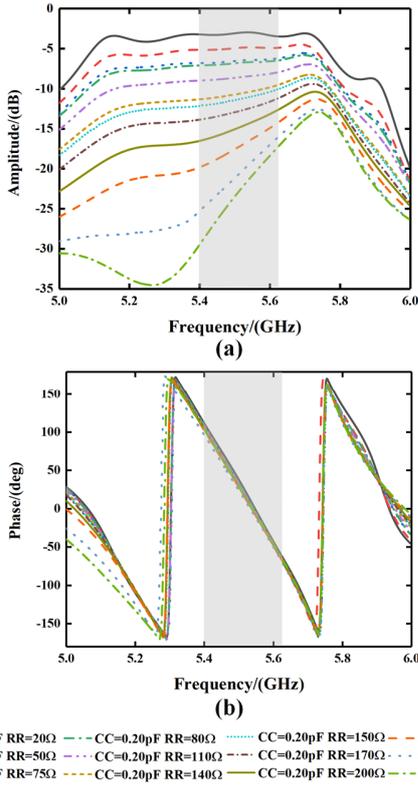

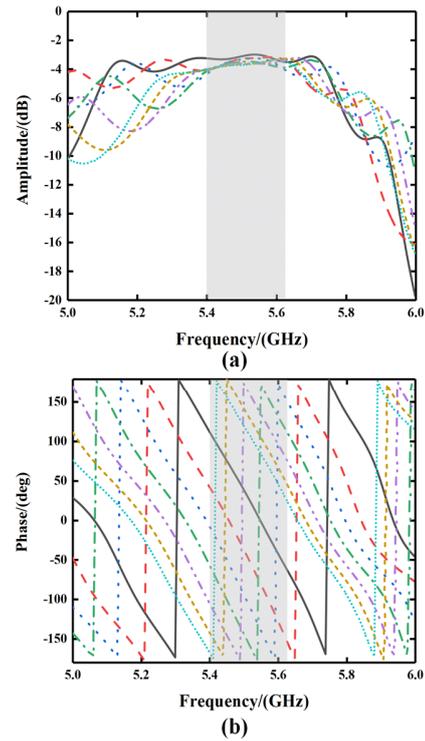

Fig. 4 Simulation results of the (a) transmission amplitude and (b) transmission phase when the equivalent resistance RR of PIN diodes varies from 20 ohm to 280 ohm and the equivalent capacitance CC of the varactors keep constant as 0.2 pF.

Fig. 5 Simulation results of the (a) transmission amplitude and (b) transmission phase when the equivalent resistance RR of PIN diodes keeps constant as 20 ohm and the equivalent capacitance CC of the varactors varies from 0.2pF to 0.8pF.

In the measurement, when the resistance in PIN diodes is 0 ohm, the corresponding bias voltage can not be added directly due to the large current. Accordingly, the range of equivalent resistances in the PIN diodes is designed from 20 ohm to 280 ohm in the simulation while the capacitances of the varactors keep constant as 0.2 pF. The simulation results of the transmission amplitude and phase shown in Fig. 4 present the good performance of the programmable attenuator. The transmission amplitude can range from -3.1 dB to -23.4 dB and the corresponding variation of transmission phase is around 15° at the frequency of 5.5 GHz. In addition, the simulation transmission amplitude and phase results of the programmable phase shifter are shown in Fig. 5. Keeping the equivalent resistance *RR* in PIN diodes as 20 ohm, the transmission phase can achieve 290° coverage with varying the equivalent capacitance *CC* from 0.2 pF to 0.8 pF, the transmission amplitude of the unit cell can change from -3.1 dB to -3.9 dB correspondingly at the frequency of 5.5GHz.

Later, the programmable attenuator and phase shifter are simulated with the receiving and transmitting antennas by changing the equivalent resistance *RR* of the PIN diodes and the equivalent capacitance *CC* of the varactors at the same time. Considering the effect of the cascaded of attenuator, phase shifter and the receiving/transmitting antennas, the dimensions of them are slightly adjusted and the final values are presented in Table1. And for illustrating the performance of independent controls of transmission amplitude and phase more directly, the typical simulation results with selected parameters of active components are shown in Fig. 6. It shows that the transmission amplitude can have the turbulence of 0.1 with changing the phases and the transmission phase can have the turbulence of 20° with changing the amplitudes from 5.39 GHz to 5.66 GHz.

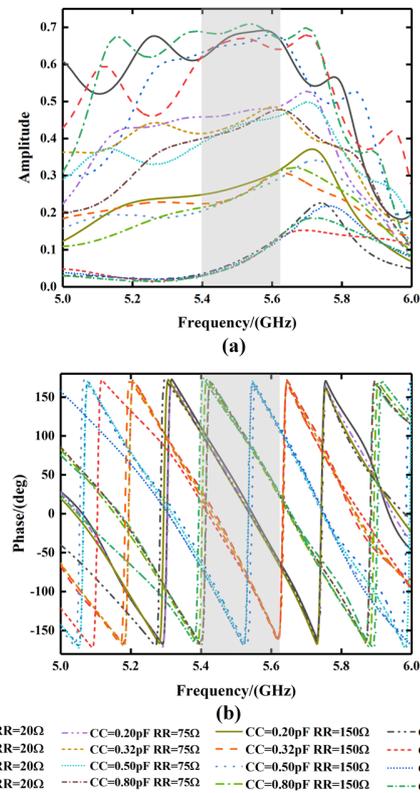

Fig. 6 The typical simulation results of the (a) transmission amplitude and (b) transmission phase when the equivalent resistance RR of PIN diodes changes and the equivalent capacitance CC of the varactors changes at the same time.

### III. ADAPTIVE BEAMFORMING AND POWER-ALLOCATION BEAMFORMING

Generally, engineering the wavefront or beamforming is the important issue in the modern active phased array systems and programmable metasurfaces such as the reflectarray and the transmitarray. With the development of the radar and communication systems, the controls of the spatial waves are not only required to radiate a single beam, but also to realize the null radiation for avoiding EM interference and the multi-beam with arbitrary power allocation. The two algorithms will be described below to lead us to complete further applications of the programmable transmitarray.

#### A. Adaptive Beamforming

The issues of adaptive beamforming [36] are to achieve the functions of EM steering beams to the desired directions with generating the null points for avoiding the interference. Some classical algorithms (e.g. the maximum signal-to-interference ratio (SIR), and minimum variance or minimum mean-square error (MSE)) have been proposed to calculate the optimal weight vector which includes the distributed information of amplitude and phase in antenna array. Correspondingly, the desired far-field patterns with main beams and null points can be obtained synthetically. Compared with the active phased array, the present transmitarray with independent controls of the amplitude and phase can achieve the adaptive beams without the complex RF chains. In our design, the maximum SIR algorithm is chosen to calculate the optimal amplitude and phase distributions for the generation of far-field pattern. The detailed description of the adaptive beamforming algorithm is introduced in the following:

The receiving signals of unit cells in transmitarray generally consist three parts of the expected signals, interference signals and the noise, which can be described as:

$$\vec{x} = \vec{x}_s + \vec{x}_i + \vec{x}_n$$

$$= \begin{bmatrix} \vec{a}_{s1} & \vec{a}_{s2} & \cdots & \vec{a}_{sN} \end{bmatrix} \begin{bmatrix} s_1 \\ s_2 \\ \vdots \\ s_N \end{bmatrix} + \begin{bmatrix} \vec{a}_{i1} & \vec{a}_{i2} & \cdots & \vec{a}_{iM} \end{bmatrix} \begin{bmatrix} i_1 \\ i_2 \\ \vdots \\ i_N \end{bmatrix} + \vec{n} \quad (1)$$

where $\vec{x}_s$, $\vec{x}_i$ and $\vec{x}_n$ represent the expected signals, interference signals and noise respectively. And the vector $\vec{a}_k$ refers to the steering vector under the signal arrival direction $\vec{\theta}_k$. Considering the weight vector, the whole receiving energy of unit cells can be derived as:

$$y = \vec{w}^H \left[ \vec{x}_s + \vec{x}_i + \vec{n} \right] = \vec{w}^H \left[ \vec{x}_s + \vec{u} \right] \quad (2)$$

where $\vec{w}$ represents the weight vector including the distributed information of the transmission amplitude and phase in the transmitarray. And $u$ represents the sum of the unexpected signals including the interference signals and noise.

By calculating the covariance matrix of the expected signals $R_{ss}$ and unexpected signals $R_{uu}$, the output power of the expected signals $\sigma_s^2$ and unexpected signals $\sigma_u^2$ can be given respectively as:

$$\sigma_s^2 = E\left[\left|\vec{w}^H \cdot \vec{x}_s\right|\right] = \vec{w}^H \cdot \vec{R}_{ss} \cdot \vec{w} \quad (3)$$

$$\sigma_u^2 = E\left[\left|\vec{w}^H \cdot \vec{u}\right|\right] = \vec{w}^H \cdot \vec{R}_{uu} \cdot \vec{w} \quad (4)$$

In our design, the SIR selected as the optimized object, can be defined as the ratio between the output power of the expected signals and the one of the unexpected signal, that is given as

$$SIR = \frac{\sigma_s^2}{\sigma_u^2} = \frac{\vec{w}^H \cdot \vec{R}_{ss} \cdot \vec{w}}{\vec{w}^H \cdot \vec{R}_{uu} \cdot \vec{w}} \quad (5)$$

According to the Eq. (5), the required transmission amplitude and phase distribution can be acquired by calculating the optimal weight vector $\vec{\omega}$ for achieving the maximum SIR. And the vector $\vec{\omega}$ can be obtained by solving the eigenvector equation (given as Eq. (6)) which is the solution when the derivative of SIR is set to zero.

$$\vec{R}_{uu}^{-1} \vec{R}_{ss} \cdot \vec{w} = SIR \cdot \vec{w} \quad (6)$$

Based on the method of the adaptive beamforming, a line array composed of 8 unit cells is analyzed in the *MATLAB 2018.0*. By predesigning the interference signal direction at 5° and the expected directions of incoming signals at -15° and 35°, the amplitude and phase distributions of the transmitarray can be obtained with the proposed adaptive beamforming algorithm. Correspondingly, the desired result of far-field pattern under the calculation followed by *SIR* method is represented by black solid line as shown in Fig. 7. However, due to the calculated results based on SIR method which is optimal rather than analytical, it can cause the difference of the two main beams more or less. And the full coverage of the transmission amplitude and phase can not be achieved ideally based on the design of physical textures. Therefore, the result of radiation pattern synthesized by the distribution of amplitude and phase of unit cells acquired in simulation is represented as red dotted line in Fig. 7. In addition, the one of radiation pattern synthesized only by the distribution of and phase of unit cells acquired in simulation is also represented as blue dashed line in Fig. 7. Based on the compared results, it can be observed that the information of transmission amplitude is very important to complete the issue of adaptive beamforming.

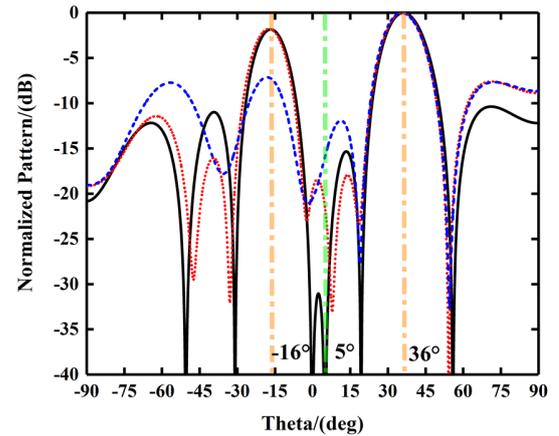

Fig. 7 Simulation results of adaptive beamforming algorithm in MATLAB. The expected signals are predesigned at -15º and 35º and the interference signal direction is predesigned at 5º. The black solid line presents calculated result of radiation pattern synthesized with the distribution of magnitude and phase acquired under the SIR method. The red dotted line refers to the radiation pattern synthesized with the distribution of magnitude and phase of unit cells in full-wave simulation. The blue dashed line presents the radiation pattern synthesized only with the distribution of phase of unit cells in full-wave simulation.

### B. Power-Allocation Beamforming

The design of multi-beam antenna using metasurfaces is very attractive in advanced electrical system. Although the radiated directions of beams can be achieved due to the previous methods [43, 44], the power allocation for multi-beams cannot be realized due to the lack of the amplitude information.

Due to synthesized method of transmitarray, the radiation pattern in the far-field can be calculated by

$$F(\theta) = \sum_{n=1}^{N} \vec{I_n} \vec{E_n} e^{jkd(n-1)\sin\theta} \quad (7)$$

where *k* and *d* represent the wave number and the period of unit cell. And $\vec{I_n}$ refers to the incident energy of the *n-th* unit cell, and $\vec{E_n}$ indicates the transmission response composed of the magnitude and phase under the *n-th* unit cell. In the simulation and experiment, the ideal plane wave is considered as illuminated source, and $\vec{I_n}$ will be uniform accordingly. As an example, to achieve the dual beams with power allocation which function of radiation pattern is given as Eq. (8),

$$F(\theta) = F_1(\theta) + \eta F_2(\theta) \quad (8)$$

where $\eta$ indicate the power ratio of the two radiation beams of $F_1(\theta)$ and $F_2(\theta)$ respectively, the transmission response of the unit cell should be given as

$$\vec{E_n} = \frac{1}{\sqrt{1+\eta^2}} \left( e^{j\varphi_{n1}} + \eta e^{j\varphi_{n2}} \right) \quad (9)$$

where $\varphi_{n1}$ and $\varphi_{n2}$ represent the required phase compensations for each beam design seperately.

To verify the theoretical method for realizing the power allocation of dual beams, the radiation directions of -15° and 30° are predesigned seperately and the power ratio of them is set as 3 dB. According to the Eq. 9, the required transmission response of unit cells in transmitarray can be obtained, and the synthesized method is applied to calculate the radiation pattern in the far field. In detail, the one-dimensional transmitarray composed of 8 ideal unit cells is considered in Matlab simulation and the corresponding results represented as black solid line are shown in Fig. 8. The performance which is coincident with theoretical analysis for generating the dual-beam power allocation is observed. And the synthesized work, based on the unit cells with amplitude and phase information acquired in our full-wave simulation, is represented as the red dotted line. And the calculation results have a little derivation from the theoretical analysis that is caused by using of the unit cells with incomplete coverage of amplitude and phase. In addition, the result obtained by the synthesized method only using the unit cells of phase information in full-wave simulation is indicated as blue dashed line, and it can also notify the importance of the amplitude information for this issue.

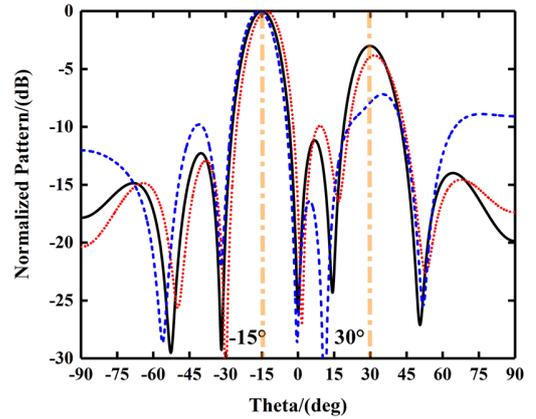

Fig. 8 Simulation results of power allocation algorithm in MATLAB. The two beams are predesigned at the directions of -15°and 30°, and the beam at the 30° is predesigned 3db lower than the beam at the direction of -15°. The black solid line presents calculated result of radiation pattern synthesized with the distribution of magnitude and phase acquired under the power allocation method. The red dotted line refers to the radiation pattern synthesized with the distribution of magnitude and phase of unit cells in full-wave simulation. The blue dashed line presents the radiation pattern synthesized only with the distribution of phase of unit cells in full-wave simulation.

## IV. EXPERIMENT RESULTS

### A. Measurement of Unit Cell in Waveguide

To verify the performance of the proposed transmitarray unit cell, a sample was fabricated and measured using two standard waveguides of WR-159, and the experimental setup [1, 45] is shown in Fig. 9. Two ad-hoc waveguide-to-unit cell transitions are designed with the thickness of 3 mm (~0.056 λ), which can convert the aperture from the size of the waveguide port to the size of unit cell by 3D printing. The device under test (DUT) is surrounded by via holes as the substrate integrated waveguide (SIW) to imitate the perfectly electric conductor (PEC) boundary. In addition, two aluminum sheets are used to keep the air distance between different dielectric layers of the proposed programmable transmitarray unit cell.

The phase can be changed with tuning the voltages on the varactors. And the four varactor diodes are controlled simultaneously with the same voltage by the DC bias lines, which are connected through 1 nH inductances as RF chock. In addition, the bias line should be isolated with slot through the boundary of the metal sheet. Similar to the design of the programmable phase shifter, the DC bias lines are also applied in the programmable attenuator to added voltages on the PIN diodes with the same method. When the EM waves are emitted from the waveguide, it passes through the ad-hoc transition first, and returns to another waveguide by another transition after going through the unit. The transmission amplitude and phase in the vector network analyzer (VNA) are measured while continuously tuning the DC source. And the final results can be acquired after calibrating the waveguides with Through-Reflect -Line (TRL) method.

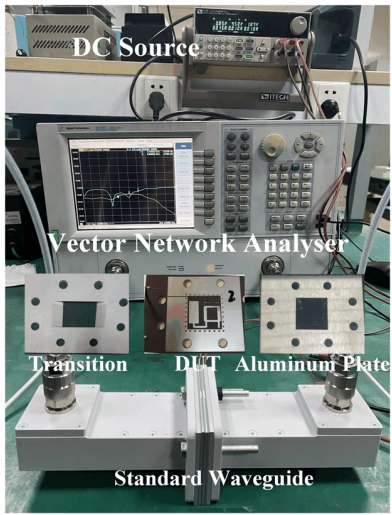

Fig. 9 The experimental setup of the standard waveguide measurement and devices under test (DUT).

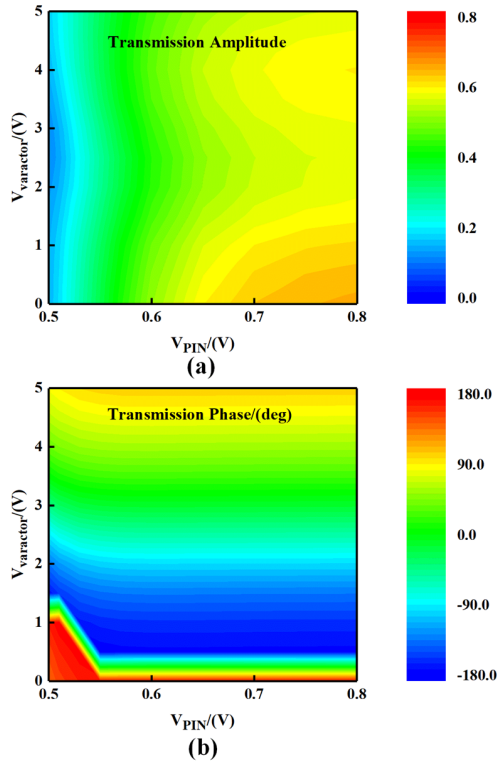

Fig. 10 The measured transmission (a) amplitudes and (b) phases of the fabricated unit cell with the variation of the bias voltages on the PIN diodes and the varactor diodes in the waveguide simultaneously under the frequency of 5.4 GHz.

After the calibration under measurement, the transmission amplitude and phase of unit cell are obtained and shown in Fig. 10. The results indicate that the transmission magnitude can achieve the range from -3.47 dB to -17.74dB with changing the bias voltage from 0.8V to 0.5 V on the PIN diodes. And the transmission phase can realize 290° coverage with changing the bias voltage from 0V to 5 V on varactors under the frequency of 5.4 GHz. And the transmission amplitude can have the turbulence of 0.1 with changing the phases and the transmission phase can have the turbulence of 40 ° with changing the amplitudes from 5.38 GHz to 5.45 GHz.

The measurement results can verify the ability of the independent manipulation for transmission amplitude and phase of our proposed active unit cells. However, there exist some slight differences between the waveguide measurement results and the simulation results. The reasons mainly include three aspects: the first is because the boundary is set as master and slave boundaries during simulations but PECs in measurements; the second is because the impinging EM wave to the unit is under oblique incidence in the standard waveguide; and finally, the application of transitions will also introduce errors to the experiments. The corresponding programmable transmitarray is fabricated to further verify the design of independent controls for transmission amplitude and phase.

*B. Implementation of Transmitarray*

A programmable transmitarray composed of 8× 8 unit cells with three PCB layers is fabricated. The bow-tie like patch antennas located on upper and bottom layers are absolutely consistent. And the middle layer combining with the square patch antennas, programmable attenuators and phase shifters, are shown in Fig. 11. The design of the bias lines in the transmitarray is similar to the method presented in the measurement of the unit cells. Accordingly, the 128 bias lines connecting to the each unit cell are all loaded with inductors of 1 nH to isolate the radio signals and DC signals. The voltages can be added on the unit cells using the DuPont lines inserted on the series of the pins, which are outside the transmitarry to decrease the influence of the measurement.

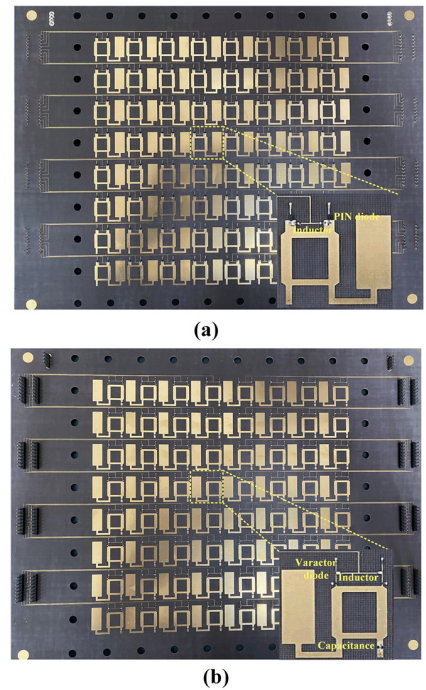

Fig. 11 The photography of the middle layer of the transmitarray. (a). The bottom of the middle layer including the receiving patch antennas and the programmable attenuators. (b) The upper of the middle layer including the transmitting patch antennas and the programmable phase shifter.

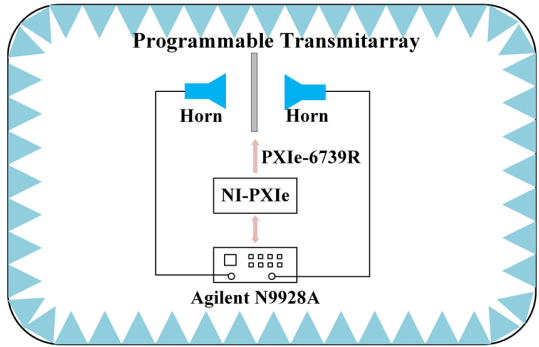

Fig. 12 Measurement setup with bias control system for calibration of programmable transmitarray.

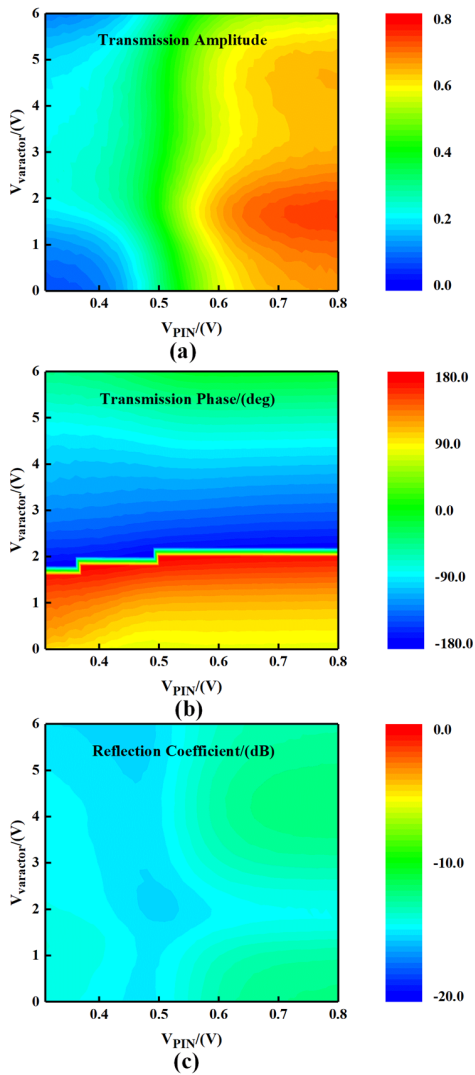

Fig. 13 The measured (a) transmission amplitudes, (b) transmission phases and (c) reflection amplitudes of the fabricated transmitarray with the variation of the bias voltages on the PIN diodes and the varactor diodes simultaneously under the frequency of 5.67 GHz.

With continuously varied bias voltages on the PIN diodes and varactors, the reflection amplitude, transmission amplitude and phase of the fabricated transmitarray are measured under two horns. The schematic of the measurement setup for calibration of proposed programmable transmitarray is shown in Fig. 12, in which, the two horns are connected to ports of the VNA (Agilent N9928A) and the bias voltages are added by the DAC module (NI PXIe-6739R) with 16-bit control (can be considered as the continuous control). With keeping all of the active unit cells be shared of the same programmable states, the transmission amplitude, transmission phase and the reflection amplitude of whole transmitarray are measured under the frequency of 5.67 GHz and the corresponding results are shown in Fig. 13. It can be observed that the transmission phase can achieve 270º coverage (the amplitude variation less than 0.1) with changing the bias voltages on varactors of phase shifters from 0V to 6 V, and the transmission amplitude can vary from -16 dB to -3.6 dB (the phase variation less than 30°) with changing the bias voltages on PIN diodes of attenuators from 0.31 V to 0.8 V in our measurement results. And the transmission amplitude can have the turbulence of 0.1 with changing the phases and the transmission phase can have the turbulence of 30° with changing the amplitudes from 5.65 GHz to 5.75 GHz. Also, the measured reflection coefficients of transmitarray are below -10 dB under all programmable states, that means the good matching work is achieved. And the measured results of the whole programmable transmitarray are consistent with the simulation results and the measured results of the unit cell in the waveguide. Correspondingly, the calibrated values of voltages loaded on the PIN diodes and varactors can be independently mapped to the responses of transmission amplitude and phase for further synthesized works of beamforming.

*C. Beamforming Results of the Programmable Transmitarray*

To verify the design of the programmable transmitarray, the fabricated sample placed on the rotating platform is measured in the far-field in microwave anechoic chamber as shown in Fig. 14. The spatial feed source for the transmitarray is selected as the conventional design of planar array composed 8× 8 microstrip patches, and the distance between the transmitarray and the patch array is about 50mm. Accordingly, the exciting port is connected to SMA connector of patch array and the other port is connected to the receiving horn located on the opposite side. And the measured data of radiation field can be sampled with rotating the platform.

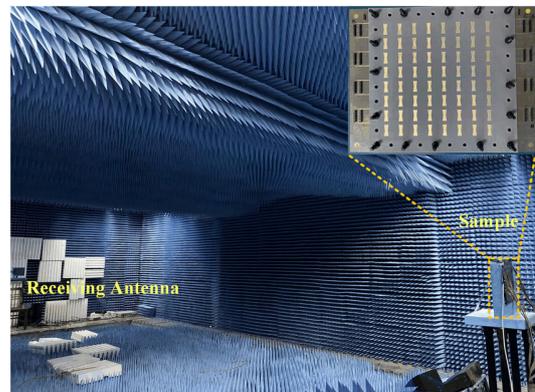

Fig. 14 The experiment setup of far-field measurement in microwave anechoic chamber.

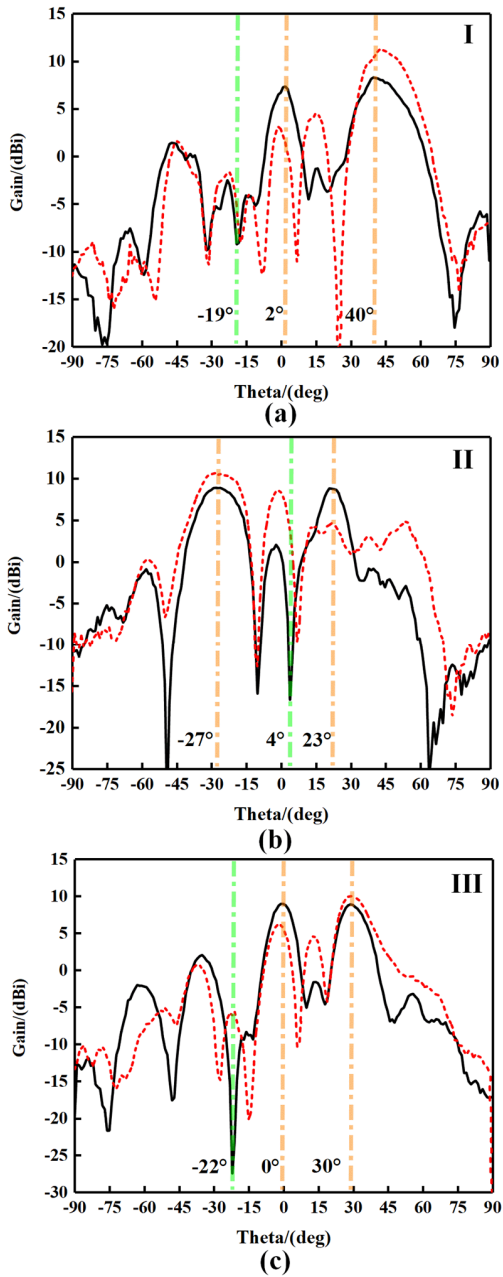

Fig. 15 Measured radiation pattern of the adaptive beamforming by using the programmable transmitarray in E-plane. The black solid lines represent the results under controls of both transmission amplitude and phase. The red dotted lines represent the ones under control of only phase.

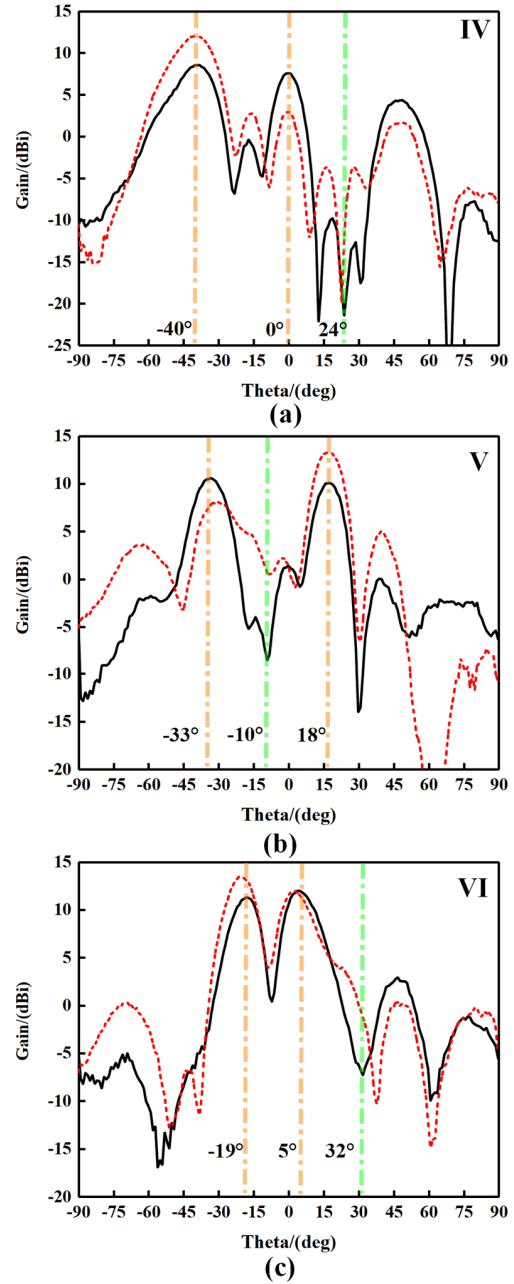

Fig. 16 Measured radiation pattern of the adaptive beamforming by using the programmable transmitarray in H-plane. The black solid lines represent the results under controls of both transmission amplitude and phase. The red dotted lines represent the ones under control of only phase.

To achieve the design of beamforming, the transmission amplitude and phase distributions should be acquired due to the method introduced in the previous part. And the corresponding bias voltages of the PIN diodes and varactors can also be obtained. Several typical beamforming works of adaptive beams and multi-beams with power allocation in both the E-plane and H-plane are measured in the microwave anechoic chamber. The predesigned angles of the expected signal and the interferences signal for the adaptive beamforming in the measurement are listed in Table 2. And the predesigned radiation angles of dual beams with power allocation ratio are shown in Table 3. Due to the complex system for programmable controls and massive bias lines, only the forward radiation which is from -90° to 90° is measured and presented.

The results of measured radiation patterns both in E and H-planes of adaptive beamforming and dual beams with power allocation are shown in Figs. 15-16 and Figs. 17-18 respectively. The good agreement between the predesigned angles and measured ones can be observed at the frequency of 5.67 GHz. And also, the construction of radiation patterns with considering both of transmission amplitude and phase can have better performance than the phase only. However, there still has the slight derivation from our requirement due to the small size ($\approx 3.78\lambda \times 3.78\lambda$) of the transmitting aperture of transmitarray and inaccurate design of amplitude and phase responses. According to the experimental results, the proposed design of independent controls for transmission amplitude and phase is verified, and the strong ability for manipulating the spatial wave under the beamforming methods is also proved.

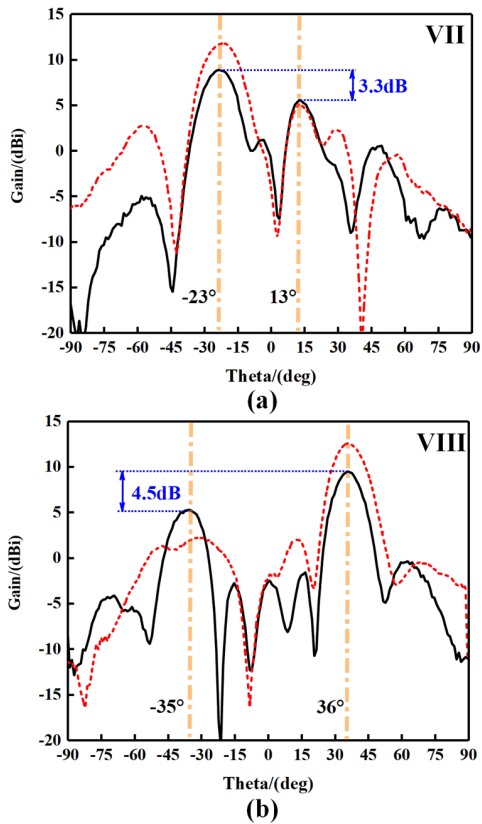

Fig. 17 Measured radiation pattern of the dual beams with power allocation by using the programmable transmitarray in E-plane. The black solid lines represent the results under controls of both transmission amplitude and phase. The red dotted lines represent the ones under control of only phase.

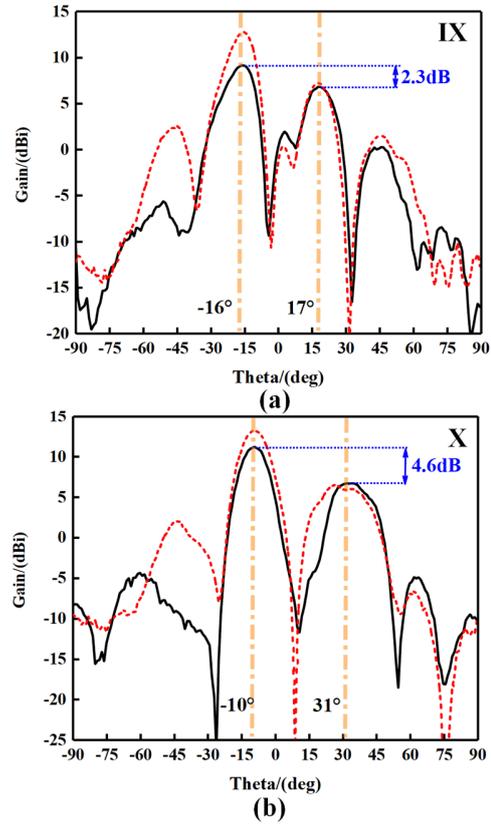

Fig. 18 Measured radiation pattern of the dual beams with power allocation by using the programmable transmitarray in H-plane. The black solid lines represent the results under controls of both transmission amplitude and phase. The red dotted lines represent the ones under control of only phase.

TABLE 2
THE PREDESIGNED DIRECTIONS OF ADAPTIVE BEAMS

| State | E/H plane | Expected Angles (deg) | Interference Angles (deg) |
|---|---|---|---|
| I | E | 0/40 | -20 |
| II | E | -30/20 | 0 |
| III | E | 0/35 | -25 |
| IV | H | 0/-40 | 20 |
| V | H | 15/-35 | -15 |
| VI | H | 5/-25 | 30 |

TABLE 3
THE PREDESIGNED DIRECTIONS OF POWER ALLOCATION BEAMS

| State | E/H plane | Beam Angles (deg) | Power Ratio (dB) |
|---|---|---|---|
| VII | E | -25/10 | -3 |
| VIII | E | 35/-35 | -5 |
| IX | H | -20/20 | -2.5 |
| X | H | -10/-30 | -5 |

V. CONCLUSION

A novel design of the programmable transmitarray that can realize independent controls of the transmission magnitude and phase is proposed. The unit cell of the transmitarray mainly consists of the cascade textures including the receiving antenna, the programmable reflection-type attenuator with PIN diodes, the programmable reflection-type phase shifter with varactor diodes and the transmitting antenna. In this paper, the unit cell is simulated and measured in the standard waveguide, and the transmitarray of 8×8 unit cells is also fabricated and calibrated. Under the 16-bit programmable control of bias voltages on PIN diodes, the transmission magnitude can range from -16 dB to -3.6 dB with the phase-error lower than 30°. And the transmission phase can range 270° with the magnitude-error lower than 0.1 when changing the bias voltage of the varactor diodes.

Accordingly, adaptive beamforming and multi-beams with required power allocation are demonstrated to verify the validity of proposed programmable transmitarray. The far-field patterns in the E-plane and H-plane are measured respectively, and both of the results show the strong abilities of beamforming under the independent manipulation of transmission amplitude and phase. And also, compared with the control of phase only in the measurement, the better results of achieving adaptive beamforming and multi-beams with power allocation are obtained by introducing the additional information of transmission amplitude. Due to the good performances of low-profile, low cost and independent controls of amplitude and phase dynamically, the proposed programmable transmitarray can have the huge potential applications in advanced radar and wireless communication systems.